\documentclass{optica-article}

\journal{opticajournal} 

\articletype{Research Article}

\usepackage{lineno}

\begin{document}

\title{Accelerated inverse design of passive \\ Si photonics}

\author{Anton Sofronov,\authormark{1,*} Dina Yakovleva,\authormark{1} Alexander Yakubovich\authormark{1}} 

\address{\authormark{1}Samsung Research}

\email{\authormark{*}a.sofronov@samsung.com} 

\begin{abstract*} 
We present an accurate differentiable parametrization scheme for topological optimization and inverse design of passive components of integrated Si photonics.
We show that the most of operations in the transformation chain from control variables to effective anisotropic dielectric tensor discretized on the Yee grid are expressible in terms of generalized convolutions that can be computed efficiently on GPU accelerators. 
Combined with the recent open-source GPU FDTD code and custom gradient-descent optimizer with dynamic control of final design manufacturability, it results in fast inverse design tool capable of producing reliable Si photonic component layouts in minutes/hours instead of days. 
The final designs, although not globally optimal, are generally physically meaningful since they reflect how fields themselves tend to propagate through the device in the most efficient way.
\end{abstract*}

\section{Introduction}
Si photonics currently is reaching the Very Large Scale Integration level \cite{shekhar2024roadmapping} driven by both recent advances in technology \cite{de2025gaas} and promising applications in sensing, communications and computing\cite{ahmed2025universal}. 
With growing complexity of Si photonic chips, there is a huge demand on design of separate functional blocks, such as passive components of intergarted photonic circuits able to provide a certain transformations of the optical fields with minimum losses and maximum efficiency. 
This is a classic problem for inverse design. 
While recent studies show some potential of AI-driven techniques for generating feasible designs from the desired functionality, it usually requires a long computationally expensive training (see recent reviews in Refs~\cite{Review2024_1, Review2024_2}).
For some specific problems the auto-differentiation or time-reversal differentiation could be a choice~\cite{Capasso2023time}.
However, passive components of Si photonics are highly suitable for adjoint method.

In this work we aim to share our experience in developing a topology-optimization-based inverse design toolchain for silicon photonics, built upon available open-source electromagnetic solver. This toolchain leverages the computational capabilities of modern GPUs and significantly speeds up inverse design process. To demonstrate its effectiveness, we present a series of simulation studies of passive devices within a standard single layer Silicon-on-Insulator integrated photonics platform. We also discuss fundamental limits of inverse design.

\section{Differentiating through a numerical solution of Maxwell equations}

Topology optimization based inverse design is an optimization problem upon a set of control variables that define the spatial distribution of materials within the design region. To optimize device performance, it is essential to evaluate how small perturbations in these control variables affect the electromagnetic response. These sensitivities are then used by the optimizer to iteratively update control variables, ultimately converging toward a design with a minimum possible deviation from the target.

A huge number of applications in Si photonics implies that device performance can be expressed in terms of frequency domain optical fields. Typical case is targeting a scattering matrix or modal transmission coefficients at the certain wavelength. In this case, the most efficient way to build a differentiable model is adjoint method together with either finite difference frequency domain (FDFD) or finite difference time domain (FDTD) solvers. Typical scenario is exciting the simulation with mode source at input port, computing resulting fields with a solver and calculating an amplitude modal coefficients $\alpha$ at the output port(s) to the desired output modes.
While FDFD works fine with toy two-dimensional models, finite difference discretization of full-vectorial 3D models on the Yee grid in frequency domain usually results in very ill-conditioned sparse system matrix, which is hard to solve even with iterative linear solvers, together with strong memory complexity. Time domain solvers can be configured to compute desired frequency domain fields by Fourier transforming field transients or by exciting the simulation with time-harmonic sources and calculating the amplitude and phase of the fields as they reach steady-state at every point of simulation domain. 

In both cases, the sensitivities of the modal transmission $|\alpha|^2$ with respect to small variation of the device shape can be calculated with only one additional adjoint simulation. 
As it is widely discussed elsewhere~\cite{Hansen:15, molesky2018inverse}, for a given parametrization $\hat\varepsilon(\vec{r}; C)$ of spatial distribution of the dielectric constant tensor $\hat\varepsilon$ within designable region $\vec{r} \in \Omega_D$ with respect to control variables $C$, the derivative $d|\alpha|^2/dC$ is given by
\begin{equation}
  \dfrac{d|\alpha|^2}{dC} = - 2\Re\left( \dfrac{d \hat\varepsilon}{dC} \cdot (\vec{E}_F \vec{E}_A)\Big|_{\Omega_D} \right),  
  \label{eq_derivative}
\end{equation}
where $\vec{E}_F$ and $\vec{E}_A$ are forward and adjoint fields, respectively.
\footnote{
\label{footnote_jac}
Note, that Eq.~(\ref{eq_derivative}) makes sense for fields and dielectric constant that are discretized at some finite difference grid of the size $N_x\times N_y\times N_z$ . In this case, $\vec{E}$ is 4-dimensional array of complex numbers, with first dimension stands for cartesian field component (i.e. $E_x$, $E_y$, and $E_z$), and remaining three dimensions are for spatial coordinates. Fields product in the inner brackets of Eq.~(\ref{eq_derivative}) is element-wise product of forward and adjoint fields.  With zero off-diagonal components of material tensor $\hat\varepsilon$ (see below), it is represented by the same-shaped 4-dimensional array with first dimension for storing $\varepsilon_{xx}$, $\varepsilon_{yy}$, and $\varepsilon_{zz}$ components. In this sense, $\hat\varepsilon$ and $\vec{E}$, having the same array shape, belongs to the same abstract space. The Jacobian  $\dfrac{d \hat\varepsilon}{dC}$ maps from the space of discretized fields to the space of control variables. 
With all field/tensor arrays flattened to a single vector of length $N = 3N_xN_yN_z$, and control variables $C$ flattened to a vector of length $M$, it is $N\times M$ matrix and $\cdot$ sign stands for matrix  multiplication. 
}

\begin{figure}[htbp]
\centering\includegraphics[width=\textwidth]{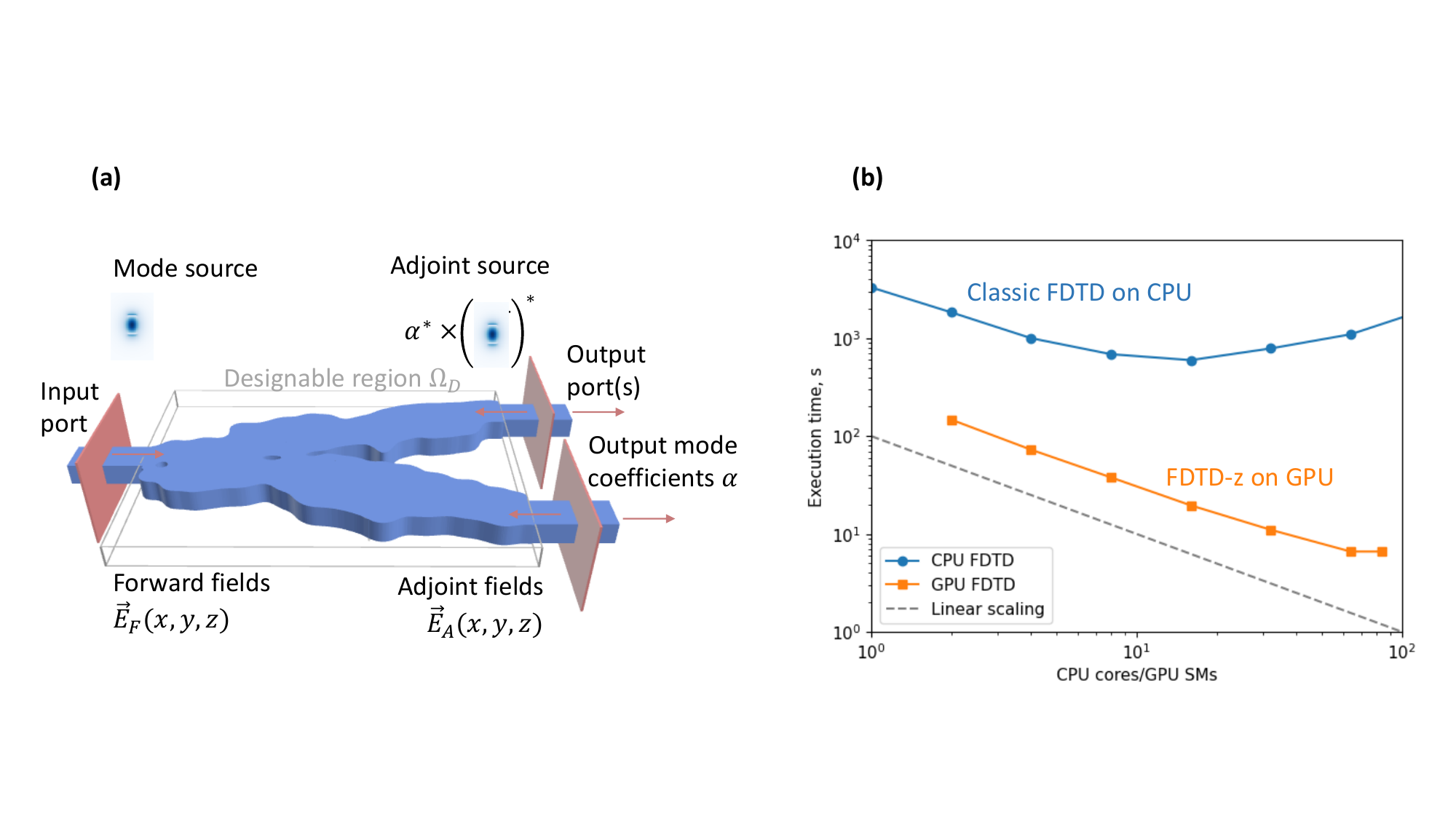}
\caption{
\label{figure1}
Schematics of forward and adjoint simulations (a) and scaling (b) of the classic FDTD algorithm on multi-core CPU and diamond FDTD algorithm on multi-SM GPU (dipole source in vacuum, 500x500x64 cells, $10^4$ time steps)
}
\end{figure}

As illustrated in Figure~\ref{figure1}(a), forward field is just a field inside design region excited by a input mode source. Adjoint field in frequency domain should be computed by solving an adjoint system excited by adjoint source, which is derivative of the target for optimization with respect to forward field, i.e. $d|\alpha|^2/d\vec{E}_F$. In time domain, adjoint system corresponds to time reversal. When time-domain simulation is excited by time-harmonic source, it is possible to replace time-reversed simulation with usual time-stepping scheme with consequent conjugating the resulting complex fields. 
It can be shown (see Appendix~\ref{App1_adjsource}) that, when targeting the modal transmission $|\alpha|^2$, exciting the simulation with adjoint source is equivalent to placing mode sources at the output ports that excite a target modes, with an amplitude coefficient depend on $\alpha^*$. 

A differentiable numerical electromagnetic solver requires only one additional adjoint simulation, but the overall performance of the inverse design process and time required for optimization is defined both by the efficiencies of solver itself and scheme for computing Jacobian $d\hat\varepsilon/dC$. 
The traditional FDTD algorithm implemented as parallel program runing on multiple CPUs suffers from the challenge of processing vast amounts of data while being constrained by limited memory bandwidth. A novel reformulation of the FDTD method~\cite{levchenko2016diamondtorre}, and open-source project fdtd-z~\cite{lu2023fdtd} that shares some of ideas of~\cite{levchenko2016diamondtorre}, minimizes data transfers between iterations by leveraging the systolic array architecture of GPUs. This approach enables fdtd-z to compute FDTD update equations with almost maximum possible simulation throughput.
This approach employs a specific “rotation” of the computational x-y-z-t space, simplifying the algorithm’s dependency graph and localizing most computations within small diamond-like cells, which are efficiently mapped to individual GPU threads. 
It allows an efficient parallelization of the time-stepping code between multiple GPU streaming multiprocessors (SM).
As we plot in Figure~\ref{figure1}(b), running FDTD on GPU allows up to $\times 100$ speed-up of the simulation compared to classic FDTD code implementation on CPUs. 

In the next section we discuss the scheme of differentiable parametrization $\hat\varepsilon(C)$ of design region geometry suitable for topology optimization, which does not slow down the two-orders-of-magnitude speed-up of fdtd-z solver.

\section{Parametrization of design region}

We start with common "density-based" technique for topology optimization when design region material is treated as a fictious mix of core and cladding, such that the designable  spatial distribution of permittivity $\varepsilon = \rho \varepsilon_\text{core} + (1 - \rho)\varepsilon_\text{clad}$ is weighted average with weights $\rho = \rho(\vec{r};C)$ at every point. 
For single layer passive Si photonics components it is enough to design a two-dimensional distribution $\rho(x,y)$, which represents just a layout mask, and then extrude it along $z$ axis to the thickness of Si device layer.

\begin{figure}[htbp]
\centering\includegraphics[width=\textwidth]{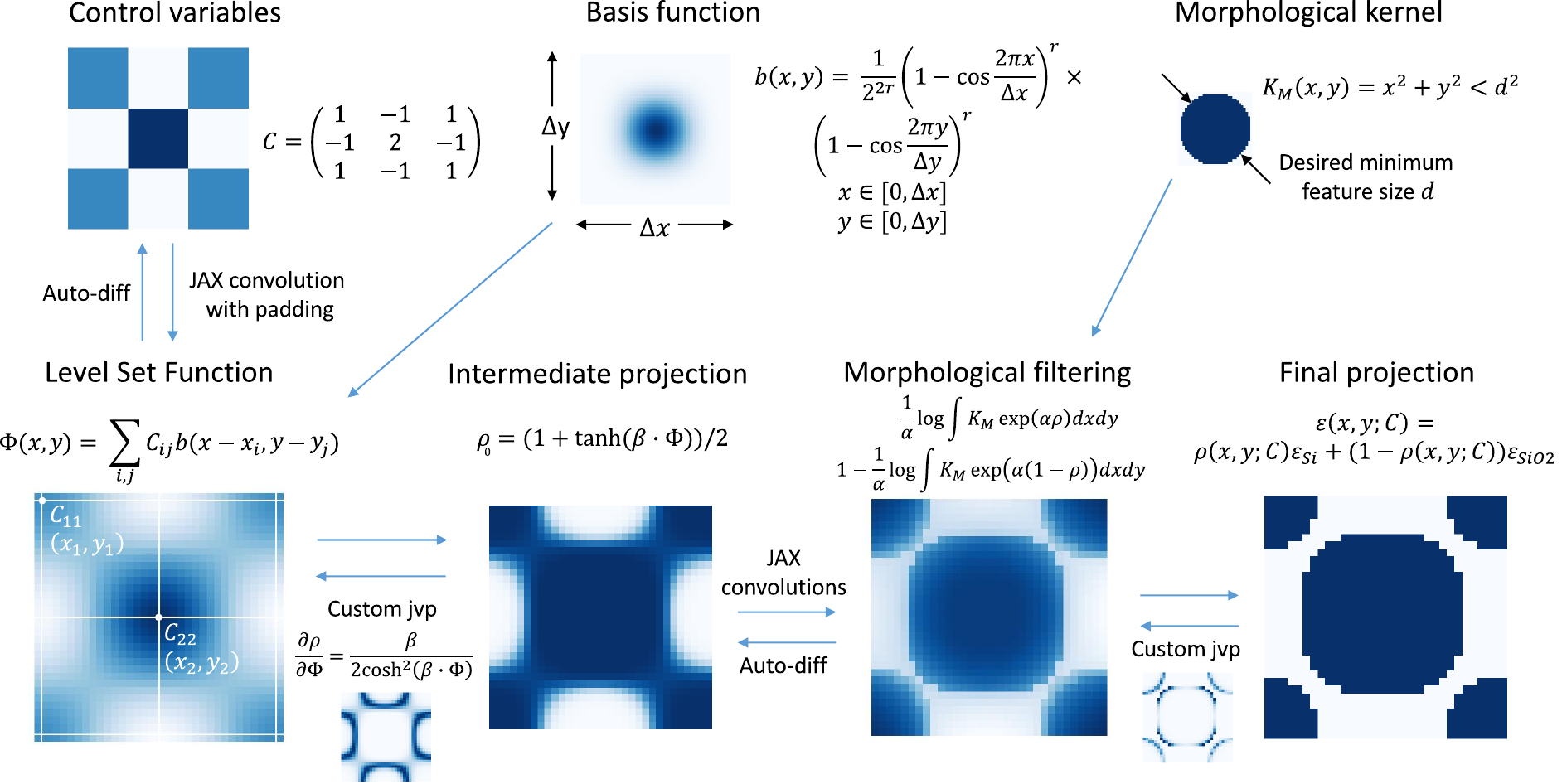}
\caption{\label{figure2}
Scheme of parametrization of design region }
\end{figure}

We adopt well-known level-set-function-based parametrization~\cite{van2013level} where the shape of the designable component (i.e. the interfaces between core and cladding materials) is formally defined by the zero-level contour of level set function (LSF) $\Phi(x,y)$, which is a  combination of compactly supported basis functions $b(x,y)$ shifted to the nodes $(x_i, y_j)$ of some rectangular grid that spans the entire designable region
\begin{equation}
    \Phi(x,y) = \sum\limits_{i,j} C_{ij} b(x - x_i, y - y_j),
\end{equation}
where control variables $C_{ij}$ parametrize the geometry. 
To keep this representation differentiable, instead of finding the contours $\Phi(x,y)=0$ implicitly, we use a hyperbolic tangent projection 
\begin{equation}
    \rho_0(x,y) = \dfrac{1}{2}\left( 1 + \tanh\left(\beta \Phi(x,y) \right) \right),
\end{equation}
such that the weights $\rho_0$ are limited in the range of $[0,1]$, and parameter  $\beta$ controls the projection strength. With very hight $\beta \rightarrow \infty$, weights $\rho_0$ takes only two values, either 0 or 1, resulting in binary design mask that corresponds to the real case when material at every point of designable region is either core or cladding. 

To avoid appearance of a topological features in the design mask that are too small to be fabricated with standard lithography processes, we incorporate a morphological filters~\cite{hammond2021photonic, schubert2022inverse} as a sequence of combination of dilate 

\begin{equation}
    \tilde \rho(x,y) = \dfrac{1}{\alpha}\log \int K_m(x-x', y-y')\exp(\alpha\rho(x',y'))dx'dy' 
\end{equation}
and erode 
\begin{equation}
    \tilde \rho(x,y) = 1 - \dfrac{1}{\alpha}\log \int K_m(x-x', y-y')\exp(\alpha(1-\rho(x',y')))dx'dy',
\end{equation}
and erode and dilate operations~\cite{sigmund2007morphology} with morphological kernel $K_m(x,y) = 1$ when $x,y$ are inside a circle with diameter $d$, representing the desired minimum feature size, and zero otherwise. 
The final weights $\rho$ are then given by subsequent projection: 
\begin{equation}
    \rho = \dfrac{1}{2} \left( 1 + \tanh\left(\beta \left(\tilde\rho - \dfrac{1}{2}\right) \right)\right).
\end{equation}
Figure~\ref{figure2} illustrates the overall chain of transformations from control variables to two-dimensional material distribution $\varepsilon(x,y;C)$.  

\begin{figure}[htbp]
\centering\includegraphics[width=\textwidth]{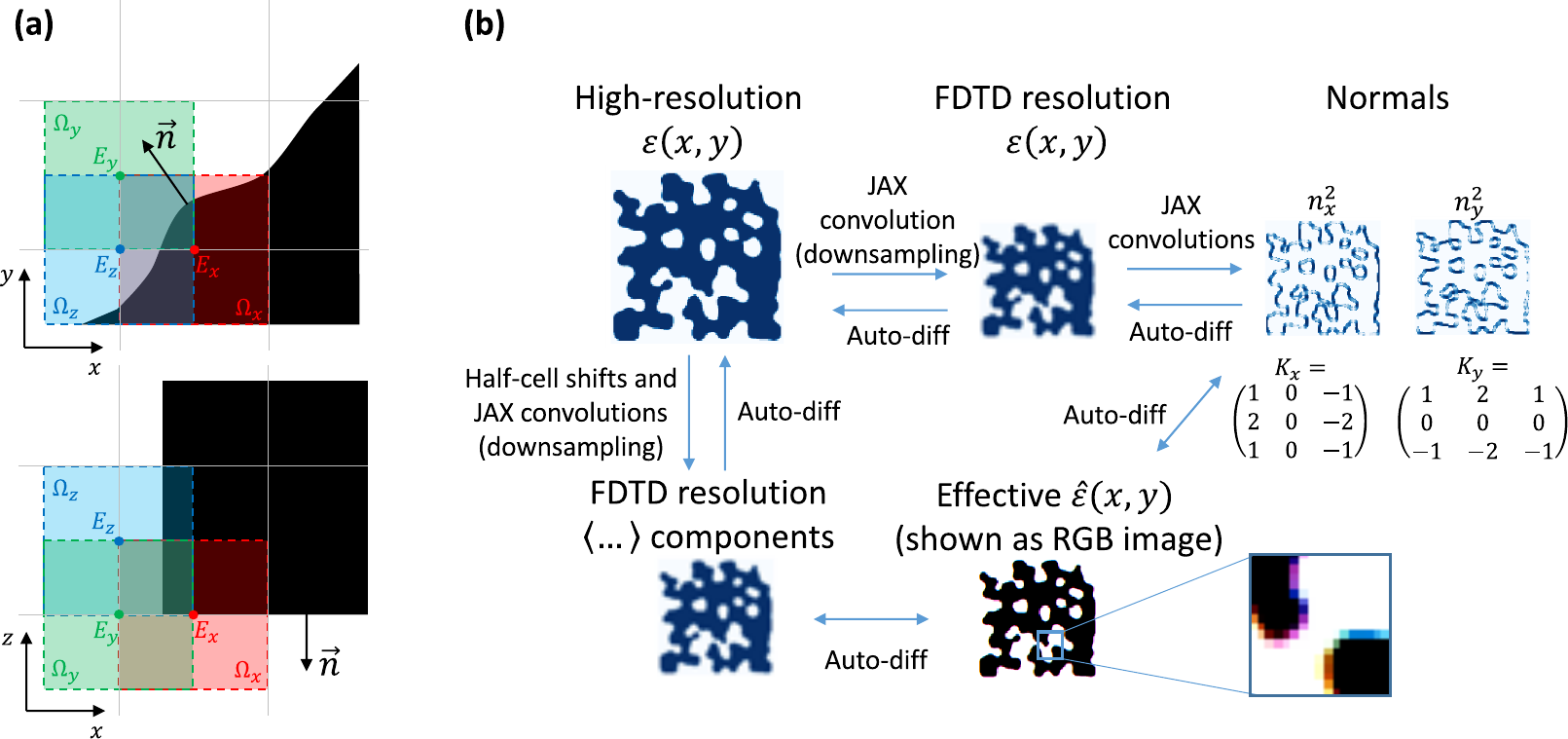}
\caption{\label{figure3}
Sub-pixel averaging: Volumes for averaging (a) and transformation chain (b) from high-resolution design mask to effective anisotropic permittivity tensor. }
\end{figure}

Nevertheless all the transformations of Fig.~\ref{figure2} are differentiable in terms of continuous $x,y$ coordinates, with discretization at any finite resolution it loses differentiability: a gradual change of any of control variables produce only spike-like changes in the pixelated binary design mask when the interface changes at least by one pixel.
That is why it is crucial to use some sub-pixel smoothing technique that assigns some effective values of the dielectric constant to the border pixels. While this effective value changes continuously with a fraction of material inside a cell, it restores the 
differentiability of the model.
We adapted an anisotropic scheme of smoothing of discontinuities on the Yee grid~\cite{Farjadpour06smoothing}, which was originally developed to improve the accuracy of FDTD when modeling discontinuous dielectric materails. 
As shown in Fig.~\ref{figure3}(a), for any Yee cell, it results in 3 diagonal components of effective permittivity tensor in the form
\begin{equation}
    \varepsilon_{ii}^{-1} = n_i^2 \langle \varepsilon^{-1} \rangle_i + (1 - n_i^2)\langle \varepsilon \rangle_i^{-1},
\end{equation}
where  $i = x$, $y$, or $z$, and $n_i$ denotes an $i^\text{th}$ component of normal vector to the interface at this cell. Averaging
\begin{equation}
\label{eq_averaging}
    \langle \dots \rangle_i = \dfrac{1}{\Omega_i} \int\limits_{\Omega_i} \dots dV
\end{equation}
is done over a volume $\Omega_i$ (of the size of the Yee cell) centered at the point where  $i^\text{th}$ component of electric field is defined.

An important observation is that the most of the transformations in Fig.~\ref{figure2} and Fig.~\ref{figure3} from control variables to sub-pixel averaged effective permitivity tensor can be expressed as generalized convolutions~\cite{dumoulin2016guide}. It opens the possibility to compute it very efficiently on GPU using highly optimized implementation in JAX project~\cite{bradbury2018jax}, which also offers auto-differentiation and just-in-time compilation of python code (note that fdtd-z project itself uses JAX to launch a cuda kernel). In practice, we use auto-differentiation almost everywhere except projection operation which is hard to auto-differentiate with high projection strengths $\beta$. For this operation we use a custom derivative rule with analytical expression for derivative of hyperbolic tangent. 

We replace integrals in Eq.~(\ref{eq_averaging}) with finite sums and use a design mask discretized at higher resolution. Then each of the $\langle\dots\rangle_x$ and $\langle\dots\rangle_y$ integrals are computed by shifting the design mask by a half-cell (on the FDTD grid) and down-sampling it to the original resolution of FDTD grid. Normal vectors at every cells are computed by convolving the design (at FDTD resolution) with the Sobel operators corresponding to numerical edge detection, with a proper normalization (Fig.~\ref{figure3}(b)).

As a result, using the JAX functionality, we get a material distribution discretized on the Yee grid, parametrized by control variables and projection strength
 $\hat\varepsilon(C, \beta)$ and ready to use as an input of fdtd-z solver, together with a function to compute a vector-Jacobian product $\left( \dfrac{d\hat\varepsilon}{dC} \cdot \vec{F} \right)$ with arbitrary field $\vec{F}$, exactly at the form of Eq.~\ref{eq_derivative}, without an implicit formation of the Jacobian itself.

\section{Optimizer with dynamic control of binarity}

The exact optimization problem to solve can be formulated as minimization of the squared deviations of all the modal transmissions of interest from their target values $T_k$:
\begin{equation}
\label{eq_target}
\begin{matrix}
    &\text{min} \sum\limits_k \left(T_k - |\alpha_k(C)|^2 \right)^2 \\
    &\text{w.r.t. } C,  
\end{matrix}
\end{equation}
where index $k$ enumerates wavelength, input port and input mode that excites the device, and output port and output mode of interest, while $\alpha(C)$ dependence is explicitly defined by the electric field that result from a permittivity of a design region: $\alpha(C) = \alpha\left( \vec{E}_F\left( \hat\varepsilon(C)\right) \right)$. 
This problem is not convex. 
In practice, however, we found that stochastic gradient-descent method usually results in reliable final designs that represent some local optimum in the parameter space. 

We use two variations of optimizers. The first one is classic Adam (short for Adaptive Moment Estimation) algorithm~\cite{kingma2014adam}. With the gradients $G_i$ given by Eq.~\ref{eq_derivative} at the $i^\text{th}$ optimization step, a bias-corrected first $\hat M_i$ and second $\hat V_i$ momenta are calculated as a moving averages
\begin{equation}
    \begin{matrix}
        M_i = b_1 M_{i-1} + (1-b_1)G_i, \\
        V_i = b_2 V_{i-1} + (1-b_2)G_i^2, \\
        \hat M_i = M_i/\left( 1 - b_1^i \right), \\
        \hat V_i = V_i/\left( 1 - b_2^i \right),        
    \end{matrix}
\end{equation}
with $b_1$ and $b_2$ being a decay factors.
The update of control variables is then given by the ratio 
\begin{equation}
    C_i = C_{i-1} - \eta \dfrac{\hat M_i}{\sqrt{\hat V_i} + \epsilon},
\end{equation}
where $\eta$ is a hyperparameter that controls the rate of convergence, and $\epsilon$ is some small number to avoid division by zero. 

The second variation is SOAP~~\cite{vyas2024soap} optimizer based on approximation of second-order Hessian from first-order gradients and running the Adam in the rotated parameter space such that updates are performed in the direction of high curvature. 
It assumes constructing two additional matrices 
\begin{equation}
    \begin{matrix}
        L \leftarrow b_2 L + (1-b_2)GG^T, \\
        R \leftarrow b_2 R + (1-b_2)G^TG, \\ 
    \end{matrix}
\end{equation} 
and their eigenvectors $Q_L$, $Q_R$, which are then used as a new basis for Adam updates:
\begin{equation}
    \begin{matrix}
        G' \leftarrow Q_L^T G Q_R, \\
        M' \leftarrow Q_L^T M Q_R, \\ 
        V'  \leftarrow b_2 V' + (1-b_2)(G')^2, \\
        N' \leftarrow \dfrac{\hat M'}{\sqrt{\hat V'}+\epsilon}, \\
        N \leftarrow Q_L N' Q_R^T, \\
        C \leftarrow C - \eta N.
    \end{matrix}
\end{equation} 

While both of these algorithms provide a good convergence in terms of control variables $C$, they do not guarantee a convergence to a true binary design. Basically, it is possible to add additional terms to Eq.~(\ref{eq_target}) that represent "unbinarity" $u(\beta)$  (relative number of gray pixels) of the design and try to minimize it with respect to $\beta$ since $\hat\varepsilon(C,\beta)$ is differentiable with respect to $\beta$ too.
Practically, we find it difficult to balance the weights of inhomogeneous optical and structural terms out. 
Instead, we complement the optimizer algorithms with specific update rule for projection strength $\beta$. It preserves the overall exponential increase with the iteration number $i$ as 
\begin{equation}
    \beta_i = \beta_\text{min} + (\beta_\text{max} - \beta_\text{min})\dfrac {\exp\left( (i/N_i)^2 \right) - 1}{e - 1 }, 
\end{equation}
where $N_i$ is the total number of iterations, but the maximum value $\beta_\text{max}$ is adjusted dynamically based of current value of $u$, again with a moving average with a decay factor $\gamma$:
\begin{equation}
    \begin{matrix}
        t \leftarrow \gamma t + (1 - \gamma)(\dfrac{1}{u} - 1)\dfrac{1}{\beta}, \\
        \beta_\text{max} \leftarrow \gamma \beta_\text{max} + (1 - \gamma)(\dfrac{1}{u_0} - 1)\dfrac{1}{t},
    \end{matrix}
\end{equation}
where $u_0$ is the desired relative number of gray pixels that is now guaranteed to reach at the final optimization step. 

\begin{figure}[htbp]
\centering\includegraphics[width=\textwidth]{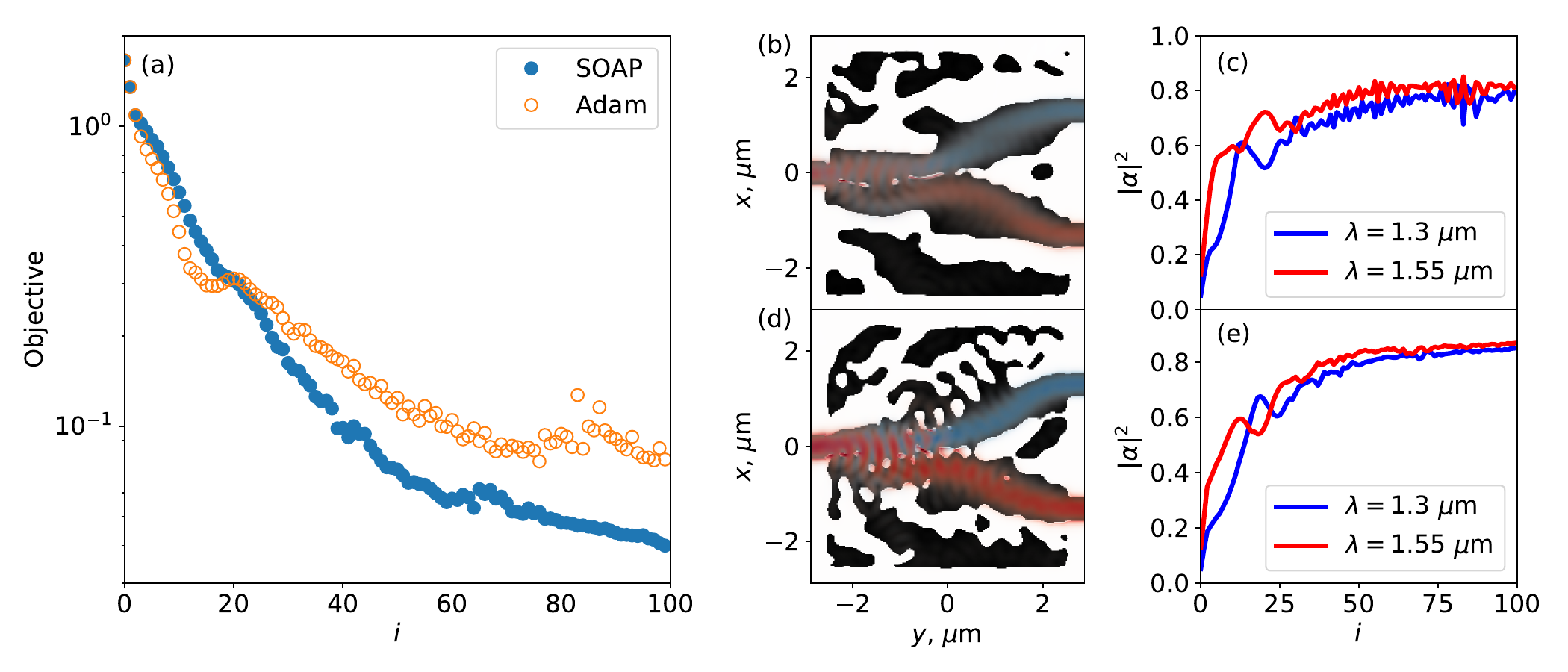}
\caption{\label{optimizers_comp}
Comparison of classic adam and SOAP-inspired optimizers:  convergence history (a), final optimized designs with magnitude of electric fields superimposed, and channel modal transmissions for adam (b, c) and SOAP (d, e) optimizers, respectively.  Morphological filtering turned off. Model size is $430\times450\times64$ cells with $dx = dy = 0.025$~nm and $dz = 0.055$~nm, design region is 210x210 cells ($5.25\times5.25$~$\mu$m) parametrized by $36\times36$ control variables. Total optimization time for 100 iteration is 1700~sec for adam and 1730 sec for SOAP optimizers ($\approx 29$~min).
  }
\end{figure}

Figure~\ref{optimizers_comp} presents a comparison of convergence of two optimizers, Adam and SOAP, with the typical example of inverse design of simple passive component that distributes $\lambda = 1.3$ $\mu$m light from the input waveguide to the first output and $\lambda = 1.55$ $\mu$m light to the second output. 
As it shown in Figure~\ref{optimizers_comp}(a), pseudo-second-order SOAP demonstrates more smooth convergence history and results in design with slightly better performance. Both simulations were started with the same initial conditions (all $C$ are zeros), with the same hyperparameters (see Appendix~\ref{App_details}) and morphological filters turned off. 
Total time required for inverse design of the component with 100 optimizer steps and full 3D electromagnetic simulation at each step is only 29 min. 

\begin{figure}[htbp]
\centering\includegraphics[width=\textwidth]{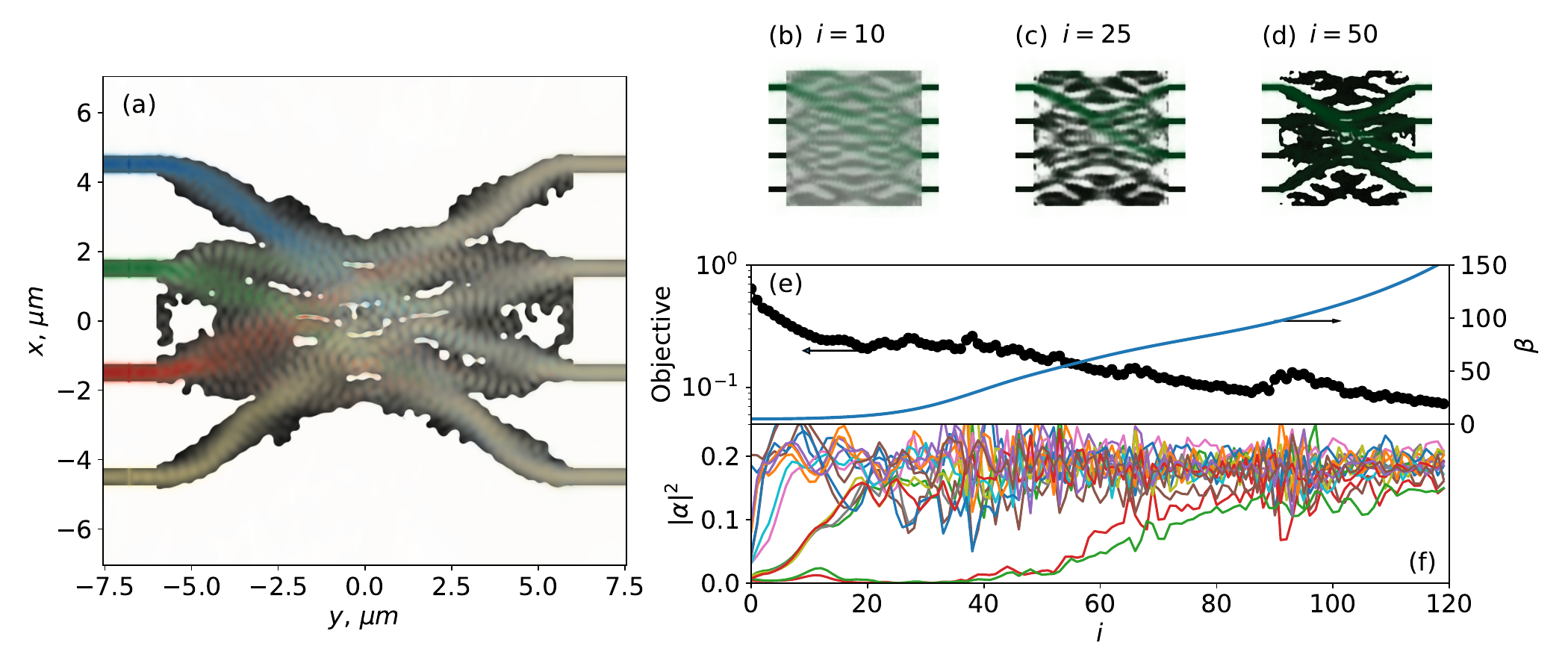}
\caption{\label{figure_4Ch_example}
Inverse design of $4\times4$ power distributor: (a) final optimized design with  fields resulting from excitation of all input channels superimposed, (b, c, d) intermediate gray-scale designs, (e) optimization history together with evolution of projection parameter $\beta$, and (f) channel modal transmissions. Green fields plotted over designs in panels (b-d)  represent the product of forward (excitation of first input port) and adjoint optical fields in logscale.}
\end{figure}

Another one example of inverse design is shown in Fig.~\ref{figure_4Ch_example}. 
It is more complex case involving a 4×4 optical power splitter designed for uniform distribution of the light from each input to the outputs. Each of the four input waveguides receives optical signals in quasi-TE mode, each with a distinct wavelength (1.52, 1.54, 1.56, and 1.58 $\mu$m respectively). The component is engineered to evenly distribute the power of each input signal to the same quasi-TE modes across all four output ports. As a result, the optimization problem consist of targeting 16 modal transmissions simultaneously. 

The final optimized design layout is shown in Fig.~\ref{figure_4Ch_example}(a).
In this case, we additionally use a "clean-up" feature when those of control variables, which corresponds to the areas where optical fields are close to zero, are gradually forced to minimal values starting from some iteration when the design unbinartity is less then 0.1. 
Total time for optimization of this component is about 9~hours since currently there is no parallelization between independent channels implemented, and all 16 modal transmissions were modeled sequentially. 

It is interesting to note how the algorithm works in physical sense. 
With $C = 0$ the initial design represents a block of material with a uniform permittivity exactly equal to the average of core and cladding between input and output waveguides. 
Both forward and adjoint simulations give some natural distribution of optical fields inside this block. The product of forward and adjoint fields (see Eq.~\ref{eq_derivative}) in this case represents some path which optical fields want to follow naturally to reach the opposite ports when excited both from the input and output side.
In this sense, it corresponds to the Lorentz reciprocity of any linear passive device with time-independent permittivity, which must have a symmetric scattering matrix, i.e. it should behave the same way if it is excited backward.
During the next iterations, the optimizer increases slightly the refractive index in the volume where both optical fields tend to localize simultaneously, while decrease it slightly everywhere else. It corresponds to supporting this natural path optical fields want to follow by confining the fields inside a higher refractive index region. Snapshots of the intermediate designs and corresponding product of forward and adjoint fields are presented in Figure~\ref{figure_4Ch_example}(b-d) to illustrate the process.
Using moving averages for both gradients and projection strength makes this process quite smooth, gradually improving the confinement and the overall device performance. 

\section{Discussion}

In this section we would like to try to address two closely related questions. The first one is very practical and can be formulated as: what area of design region is sufficient for a given functionality of inversely designed component? 
The second one is more fundamental: what is the best performance that can be achieved by inverse design with a given design area?
Both these questions are about physical limits of the photonic components. Although they have been recently addressed in part, there is no good universal answer due to intrinsic non-convexity of the problem. 

We will use a representative example of inversely designed mode converter within a discussion in this section. The functionality of the component is to convert a quasi-TE mode of $0.5\times0.22$~$\mu$m Si waveguide at the input port to quasi-TM mode at the output, while suppressing the transmission to the output quasi-TE mode as much as possible. 
Note that it should not be confused with optical isolator since the system is reciprocal~\cite{Fan2012nonreciprocal, jalas2013and}.

\begin{figure}[htbp]
\centering\includegraphics[width=\textwidth]{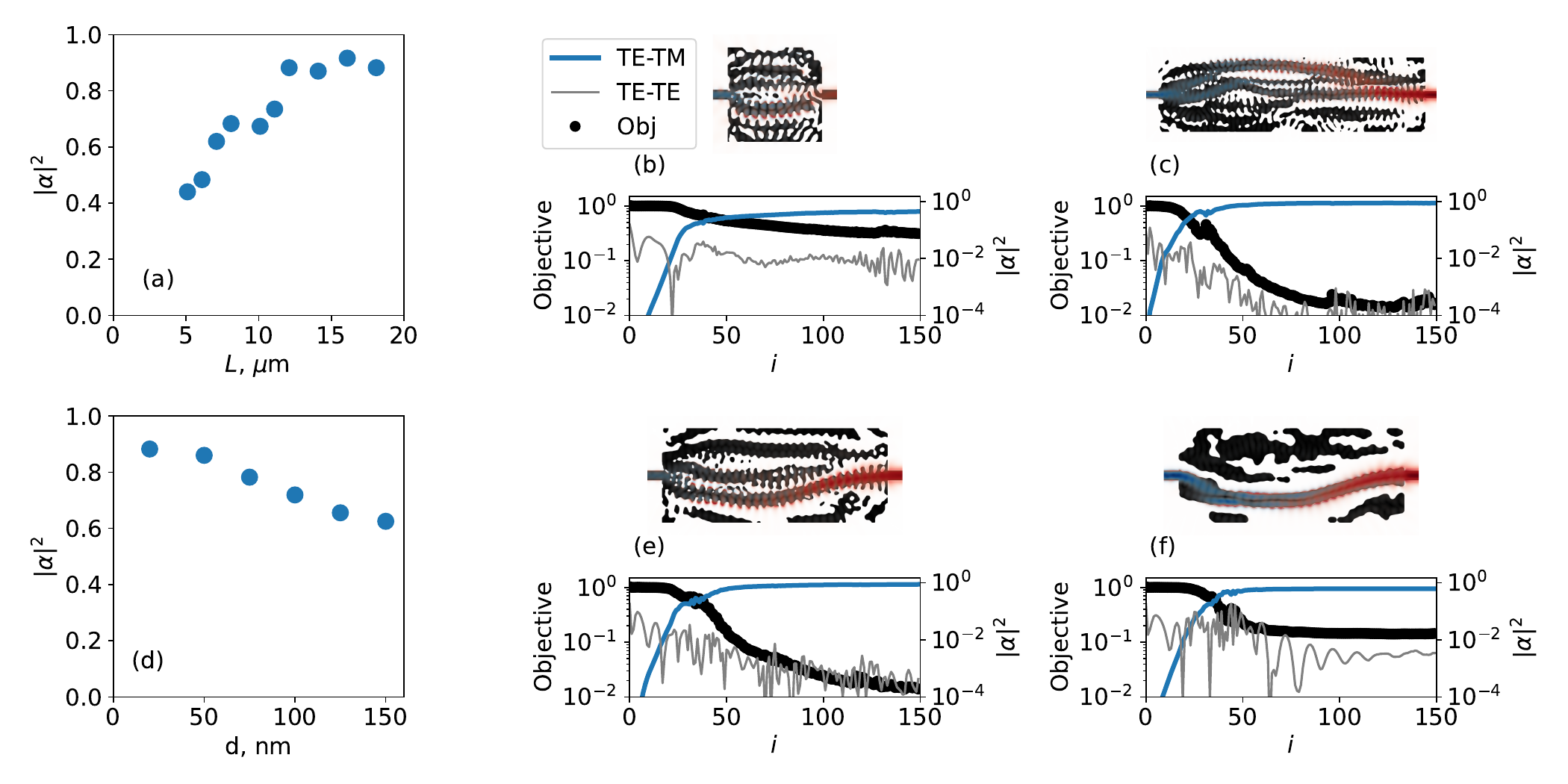}
\caption{\label{figure_mc}
Scaling of TE-TM mode converter performance. 
Drop of mode conversion with decrease of design region length (a), designs and convergence history of mode converters with $L = 5$~$\mu$m (b) and $L = 18$~$\mu$m (c) with minimal feature size set to 20~nm. Drop of mode conversion with increase of radius of morphological filter (d), designs and convergence history of mode converters with $d = 20$~nm (e) and $d = 150$~nm (f) with device length set to 12~$\mu$m. Width of design region is fixed to 5~$\mu$m.
Fields over designs represent absolute value of in-plane ($|E_x|$,  blue color) and out-of-plane ($|E_z|$, red color) component of electric field.}
\end{figure}

We perform two series of inverse designs of the mode converter with the decrease of design region length from $L = 18$~$\mu$m to $L = 5$~$\mu$m (keeping it width equal to 5~$\mu$m) and the increase of diameter of morphological kernel from $d = 20$~nm to $d = 150$~nm. Figure~\ref{figure_mc}(a) shows scaling of the mode conversion efficiency with the decrease of length. Starting from  $L \approx 11$~$\mu$m, mode conversion demonstrates a drop from steady value of almost 0.9 with $L>11$~$\mu$m down to 0.4 for the shortest device. Figures~\ref{figure_mc}(b) and
~\ref{figure_mc}(c) show the final designs and convergence history for the shortest and longest devices in this series. 

Intuitively, increase of minimal allowable structural feature size controlled by kernel diameter $d$ limits the cardinality of the parameter space, resulting in poorer performance. Indeed, as it is plotted in Figure~\ref{figure_mc}(d), mode conversion drops down to 0.6 when $d$ reaches 150~nm. Figures~\ref{figure_mc}(e) and
~\ref{figure_mc}(f) show the final designs and convergence history for the smallest and largest kernel sizes in this simulations.

Scaling in Figs.~\ref{figure_mc}(a) and ~\ref{figure_mc}(d) represents the typical behavior of inversely designed photonics components. However, there is no simple way to estimate neither bounds for a given design area nor an area sufficient for a good performance (which is usually an input parameter of the optimization problem) without some test inverse designs. 

One of the approaches could be based on diffraction heuristics~\cite{Miller2023}. 
With component functionality presented as a device operator $D$ that connects optical fields at the input with optical fields at the output of the device, a singular value decomposition of the matrix representing $D$ at discretized grid gives a set of orthogonal modes at the input and output device planes together with a singular values of $D$. 
The latter represents an "abstract optical channels" that connect input modes to output modes such a way that the desired field transformation is achieved. 
Usually, only a limited number of singular values are large enough, and device functionality can be represented by a limited number of connections, while all others provide only small corrections.
It can be shown~\cite{Miller2023} that each of these channels requires a length of $\lambda_0/2n$, with $\lambda_0$ is a vacuum wavelength and $n$ is a maximum refractive index inside the design region.

Applying this heuristics to our case of mode converter, we found that only about 50 first singular values of corresponding device operator have a noticeable values ($>0.1$~of maximum), which corresponds to $L_0 = 11.4$~$\mu$m for the case of Si. 
While this number coincides surprisingly well with the moment of performance drop in Fig.~\ref{figure_mc}(a), it looks impossible to take into account any structural limitations (Fig.~\ref{figure_mc}d) in this approach. 

As we have already mentioned, since the optimization problem in Eq.~(\ref{eq_target}) is not convex, there is no way to find a global optimum within a reasonable time. 
However, it was shown that it can be converted to a convex problem in higher-dimensional space~\cite{Molesky2022T_limits, chao2022physical, Vuckovich2023bounds} while sparsity of the problem can be utilized to solve it more or less efficiently~~\cite{Gertler2025QCQPs}. This approach is based on switching from optimization on the structural parameter space $C$ to optimization on the space of arbitrary fields with a set of additional constraints: fields must satisfy the Maxwell equations at every point of simulation domain whereas every point of designable region is filled with either core or cladding material. Generally, any kind of limitations on structural topology, including morphological filtering, can be included in this formulation.
Solving this problem results in global optimum, i.e. it gives a fundamental limit of performance for component with given functionality, volume and a set of materials it made from, which no any real device is able to outperform. 

Unfortunately, while it works fine with two-dimensional models, expanding it to a real case of 3D models, representing any feasible component of integrated Si photonics, leads to terrible memory scaling. A number of constrains, each one being a sparse matrix similar to FDFD system matrix in terms of memory complexity, grows as an inverse resolution cubed (compared to the power of two for 2D models), which makes even a setup of 3D problem for a sparse solver of Ref.~\cite{Gertler2025QCQPs} unfeasible. 
Estimation on global bounds on mode conversion purity in 2D waveguide presented in Ref.~\cite{Vuckovich2023bounds} gives some insights, but cannot be directly transferred to a three-dimensional case. 

In conclusion, more research is needed to fully address the questions from the beginning of this section. Currently it is hard or even impossible to estimate a bounds of performance for real 3D models. 
In this conditions, fast inverse design workflow is crucial for development of passive Si photonics components. 
It allows a rapid test runs and gives some overview of the landscape of possible performance, although not globally optimal, but achievable with a given area and topological constrains. 

\bibliography{references}

\appendix

\section{Adjoint mode source}
\label{App1_adjsource}

First, note that a forward mode source is a electric displacement current density $\vec{J}^F$ defined on the plane with normal $\vec{n}$  such a way it excites optical fields corresponding to electric $\vec{e}$ and magnetic $\vec{h}$ fields of the mode, and it is given by

$$
\vec{J}^F \sim \vec{n} \times \vec{h}.
$$
With $\vec{n}||z$
$$
J_x^F \sim - h_y,
$$
$$
J_y^F \sim  h_x,
$$

Modal coefficient of arbitrary field $(\vec{E},\vec{H})$ to the mode $(\vec{e},\vec{h})$ at the plane with surface element $d\vec{S}$ is given by 

$$
\alpha = \dfrac{1}{4}\left( \dfrac{1}{N} \int \vec{E}\times\vec{h}^* \cdot d\vec{S}  +  \dfrac{1}{N^*} \int \vec{e}^*\times\vec{H} \cdot d\vec{S}\right)
$$
with normalization 
$$
N = \dfrac{1}{2}\int \vec{E} \times \vec{H}^* \cdot d\vec{S}
$$

Again, with $d\vec{S}||z$, the term with implicit dependence on $\vec{E}$ is proportional to $E_xh_y^* - E_yh_x^*$.

With a given objective $F(\alpha)$, components of adjoint source $\vec{J}^A$ are
$$
J_x^A = \dfrac{dF}{d\alpha}\dfrac{d\alpha}{dE_x},
$$
$$
J_y^A = \dfrac{dF}{d\alpha}\dfrac{d\alpha}{dE_y},
$$
where from the above expressions
$$
\dfrac{d\alpha}{dE_x} \sim h_y^*
$$
$$
\dfrac{d\alpha}{dE_y} \sim - h_x^*
$$
i.e.
$$
\vec{J}^A \sim - (\vec{J^F})^*
$$
with coefficient $dF/d\alpha$. 
For $F$ given by Eq.~(\ref{eq_target}) of main text, 
$$
F = \sum (T_k - |\alpha_k|^2)^2, 
$$

$$
\dfrac{dF}{d\alpha_k} = 2( |\alpha_k|^2 - T_k)\dfrac{d|\alpha_k|^2}{d\alpha_k} = 2( |\alpha_k|^2 - T_k)\dfrac{d\alpha_k^*\alpha_k}{d\alpha_k} = 2\alpha_k^*( |\alpha_k|^2 - T_k) 
$$

\section{Simulation hardware and default parameters of optimizers}
\label{App_details}

For all the simulations in the article we use a server with GPU NVIDIA RTX A5000, classic FDTD on CPU (Figure 1(b) of main text) is meep running in parallel through MPI on 2 x AMD EPYC 7662 64-Core Processor.

Default simulation parameters:
$$
\begin{matrix}
&b_1 = 0.9 \\
&b_2 = 0.999 \\
&\eta = 0.001 \\
&\gamma = 0.95 \\
&u_0 = 0.01   \\
&\beta_\text{min} = 10 \\
& \alpha = 20 \\
& \varepsilon_\text{core} = 3.4 \\
& \varepsilon_\text{clad} = 1.44
\end{matrix}
$$

\end{document}